\documentclass[journal]{IEEEtran}
\usepackage{amsmath}
\usepackage{graphicx}
\usepackage{nomencl}
\ifCLASSINFOpdf
\else
\fi
\begin{document}
%
\title{A Simplified Approach to Two-Port Analysis in Feedback }
%
%
%

\author{Morteza~Fayazi, Ali~Fotowat, and~Zahra~Kavehvash
\thanks{The authors are with the Electrical Engineering Department, Sharif University
of Technology, Tehran, 11365-11155, Iran (fayazi@umich.edu);  (afotowat@sharif.edu);
(kavehvash@sharif.edu);}
}

\maketitle

\begin{abstract}
In this paper, a new pedagogical approach for analyzing the negative feedback circuits is proposed. The presented approach is in fact the completed form of the well-known two-port network analysis which is the most intuitive method for teaching the negative feedback concept.  The two-port network analysis is rewritten in a more general and conceptual format. In analyzing the output series feedback, the presented analysis resolves prior shortcomings. The presented approach helps the students analyze and design all types of negative feedback circuits more intuitively.
\end{abstract}


%
\IEEEpeerreviewmaketitle

\section{Introduction}
%
%
%
%
\IEEEPARstart{T}{he} concept of negative feedback is well recognized in analysis and design of electronic circuits \cite{144656}. The basic modules of an operational amplifier which produce greater than unity gain are electronically constructed using active elements. These elements are subject to severe variations versus temperature, supply voltage, process and aging. Providing a constant gain versus the above mentioned variations of active element parameters is one of the most important reasons for employing feedback in practical circuits \cite{nla.cat-vn2245999,669729,6674723}. Distortion reduction is another outcome of feedback circuits \cite{nla.cat-vn2245999}, \cite{5}. All these advantages come at the price of gain reduction.
\par
Educating a successful electronic engineer requires providing him or her with the comprehensive understanding of the role of each part of the circuit on the overall system performance features. The very basic and most accurate method for analyzing the feedback circuits is through using the well-known KVL-KCL relations. This method will definitely yield accurate results. However, writing long and complicated KVL-KCL relations gives rise to the probability of computational mistakes and these equations will hardly enhance the students' intuition. Signal flow graph (SFG) is another well-known technique for feedback analysis \cite{661648}. It helps the student perceive the mechanism of feedback by engaging a graphical method, but, due to numerous complicated mathematical equations, it fails to give them any sense of where the equations head to. Furthermore, this method does not give any insight about the effect of feedback on decreasing or increasing different circuit parameters. The other commonly used method for handling feedback circuits is Return-Ratio \cite{144656}. This method tries to compute the loop gain using the conventional approach for analyzing electronic circuits, and then the other parameters will be computed using the calculated loop gain. Still, the computed return ratio and defined loop gain have some differences while they are supposed to be equal. This technique, despite being probably the best method for advanced designers, is a bit too complex for beginning students. Perhaps the more intuitive method of feedback analysis presented in many circuit analysis textbooks, is based on the concept of Two-Port Networks \cite{1023010}.  This analysis scheme is basically based on derivation of the closed-loop parameters from the open-loop gain (a) and the feedback factor (f) \cite{704038} and is originated from two-port networks theory \cite{nla.cat-vn2245999}. In comparison to other mentioned feedback analysis techniques, this method helps the students perceive the effect of feedback on many circuit parameters such as gain, input and output impedance. Reference \cite{102847} comes up with a generalized feedback model based on two-port theory which resolves the issue of analyzing local feedback (source/emitter degeneration). Local feedback, however, can be addressed with some simple derivation not worth using feedback model.
\begin{figure*}[t]
\includegraphics[scale=0.75]{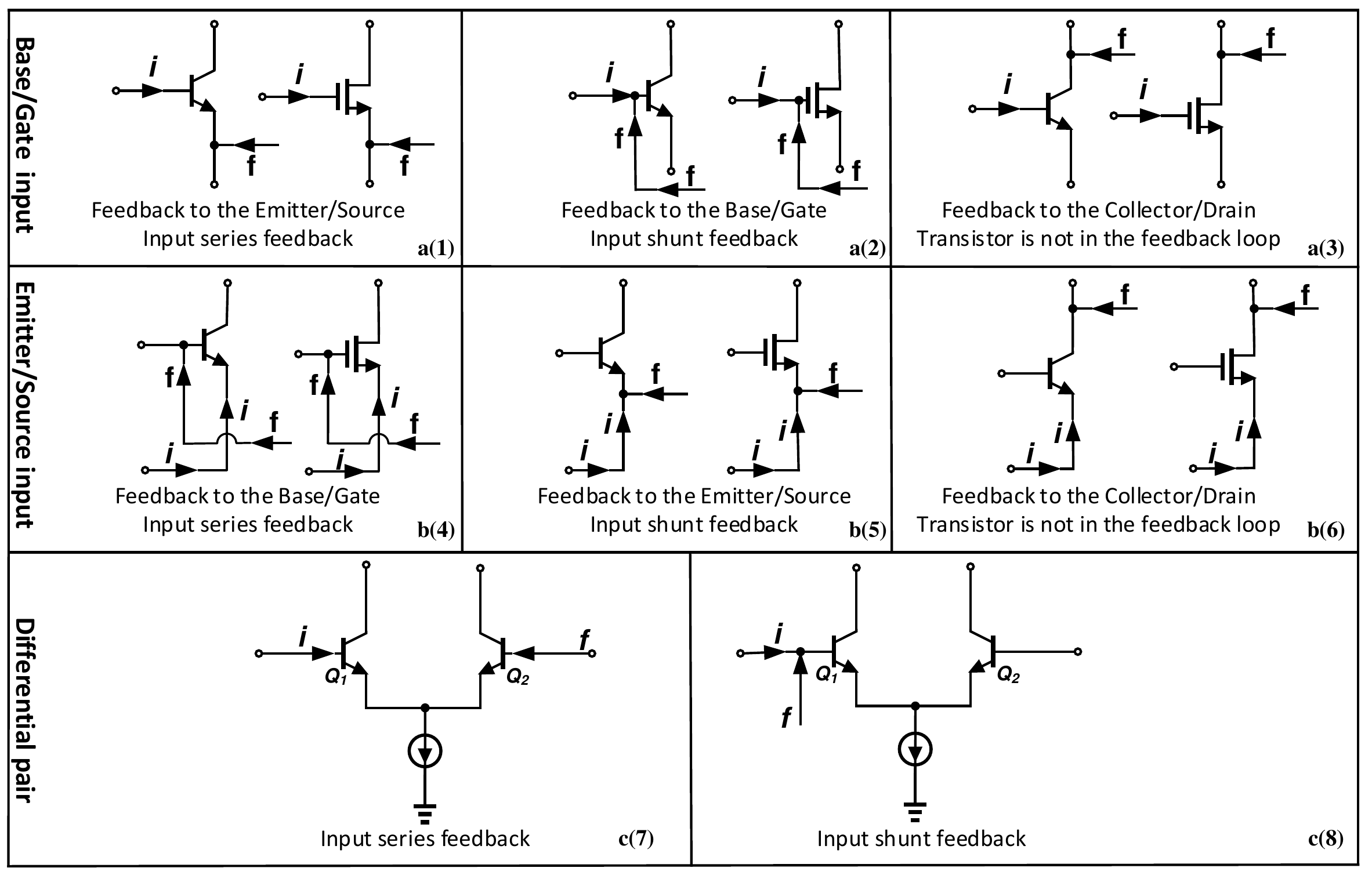}
\caption{Universal categorization of all different input feedback connections (i: input signal, f: feedback signal).}
\label{Figure. 1a}
\end{figure*}
\begin{figure*}[]
\includegraphics[scale=0.75]{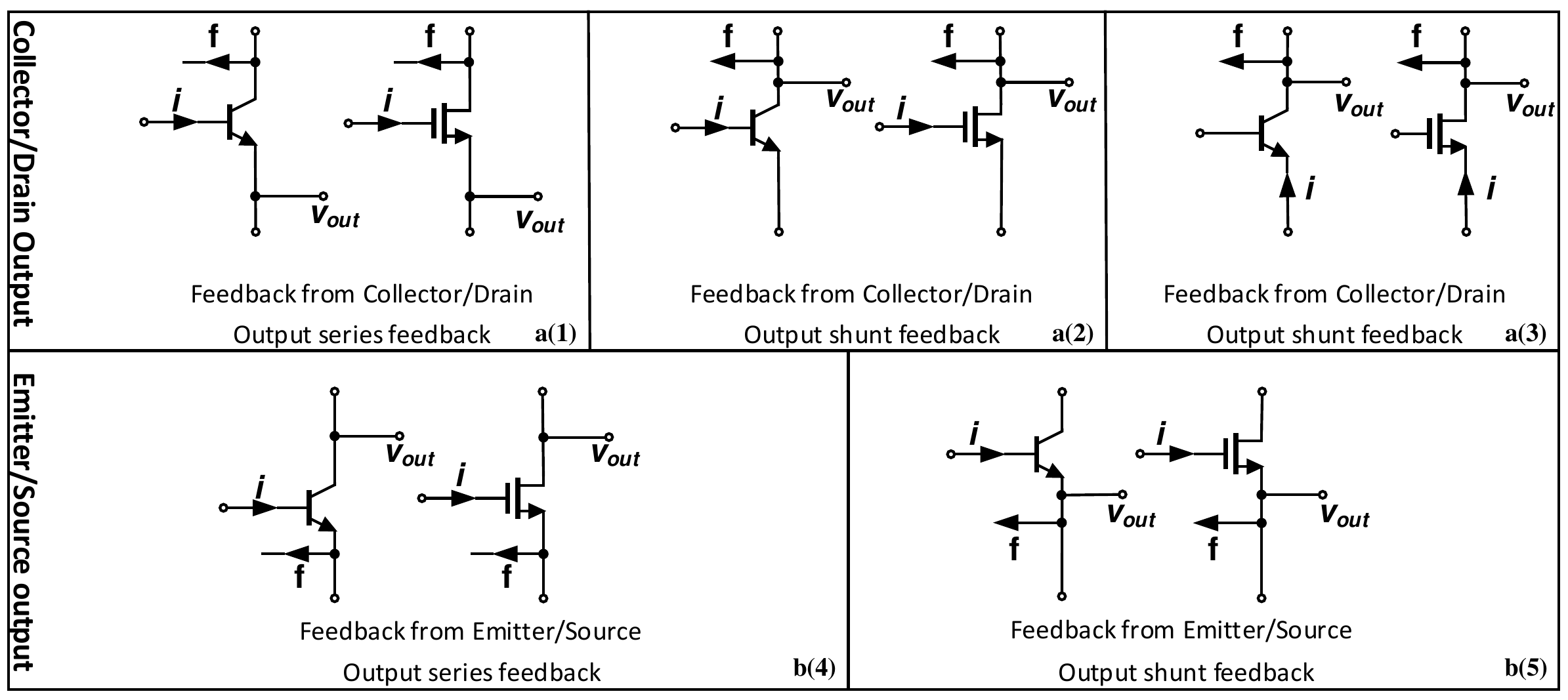}
\caption{Universal categorization of all different output feedback connections (i: input signal, f: feedback signal).}
\label{Figure. 1b}
\end{figure*}
\par
The two-port network analysis approach has its foundations in the circuit theory, but it has been developed to be specifically used in feedback circuits. It employs the concept of network matrices, such as impedance and admittance, while avoiding tedious mathematical calculations. One of the most important problems in analyzing feedback circuits through this technique is  calculation of the output impedance which is an important parameter of an operational amplifier. As an example, the main referenced books in teaching circuit analysis are unable to provide a thorough analysis of output series feedback impedance using this recent approach (the case of sensing output current) \cite{nla.cat-vn2245999}, \cite{8}. There is just a vague explanation of the output series feedback impedance in one special case which is neither complete nor entirely accurate in Gray and Meyer's book \cite{nla.cat-vn2245999} and \cite{1393103}. Another attempt for analyzing the case of output series feedback is reported in \cite{1049599}. Nevertheless, this work has failed to bring an  intuitive method while only considering a special case. A similar work has been reported in \cite{4696057} which again suffers from the same drawbacks, i.e., neither providing any intuitive understanding nor considering all cases.\par
In this paper, a complete conceptual approach based on two-port network analysis is provided for analyzing different types of feedback circuits. The output impedance is then calculated in the special case of series feedback at the output (sensing the output current). The paper is organized as follows. In section II, practical configurations and the effect of loading of four possible
feedback network configurations are investigated and analytical derivations of the output impedance of series-series feedback are then developed. We verify the presented method using KVL-KCL, as well as SFG analysis in section III. Section IV
concludes the paper. This technique is appropriate for feedback circuits with any kind of transistors. Without loss of generality, we concentrate on bipolar case since it is more general and the results could be easily extended to other circuits with MOS transistors.

\section{THE PROPOSED FEEDBACK CIRCUIT ANALYSIS}
\subsection{Practical Configurations and the Effect of Loading}
In this section, a complete conceptual approach toward the analysis of feedback circuits based on the two-port network solution is proposed.  In the two-port solution, the students must compare the circuit against one of the four possible network configurations: shunt-shunt, series-series, series-shunt and shunt-series, where loading issues can be modeled using Y, Z, h and g parameters, respectively. This process is exhaustive and needs remembering all four two-port network configurations and their relations. To make this procedure faster, easier, and more intuitive, we suggest recognizing the nature of the feedback based on the circuit configurations and feedback connections.  Fig. 1 and Fig. 2 show all possible configurations of feedback circuits when connected to the circuit input and output.
\par
The topology with feedback signal applied to the collector/drain of the output transistor is irrelevant and never used [a(3), b(6)]. This is because neither the difference between the output and input voltages nor the difference between the input and output currents are measured and fed to the circuit in this case. Furthermore, the case of a differential pair will be reduced to the case of c(7) in Fig. 1, where the feedback signal is assumed to reach Q1 through its emitter or reduced to the case c(8) in Fig. 1, where the feedback signal is assumed to reach Q1 through its base.
\par
Once the type of the feedback is recognized by matching to the cases presented in Fig. 1 and Fig. 2, which are obviously more intuitive, it remains for the student to take into account the loading effect of the feedback circuit. Referring to the four parameterized boxes (Y, Z, h and g), loading effect calculation would be cumbersome. Sedra and Smith \cite{12} use a linguistic connection relating series with open-circuit and shunt with short-circuit to help the students calculate the loading effect more easily. Instead, the loading could be calculated based on the diagrams presented here. As an example for shunt-series feedback, consider the circuit shown in case (d) of Fig. 3.
\begin{figure*}[]
\includegraphics[scale=0.7]{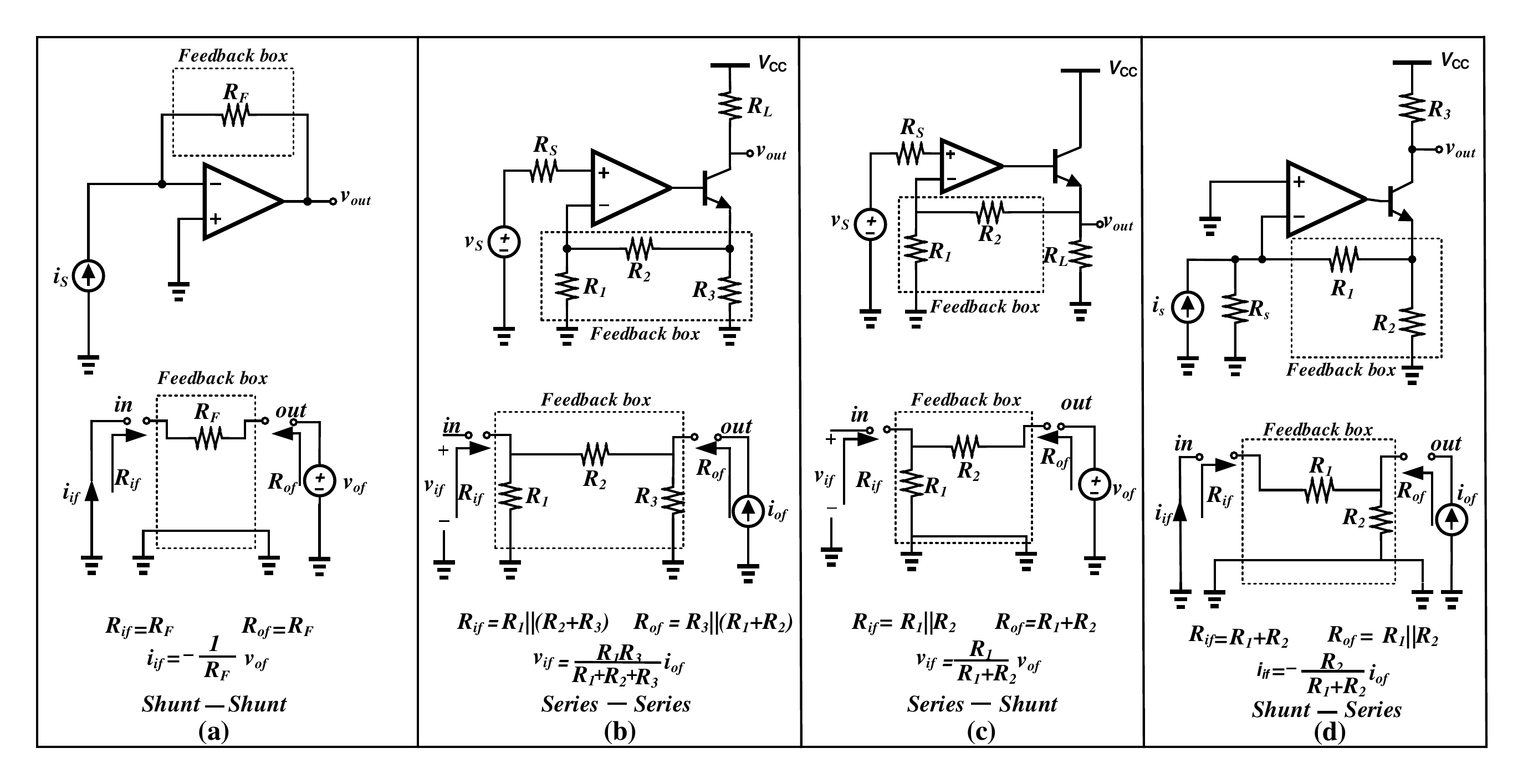}
\caption{Simple loading analysis after feedback type is recognized by using intuitively clear open/shorts $i_{of}$/$v_{of}$.}
\label{Figure. 2}
\end{figure*}
It is obvious that, $i_{if}=-(\frac{R_2}{R_1+R_2})i_{of}$ and the equivalent impedances seen through the input port, $R_{if}$, and output port, $R_{of}$, are
\begin{subequations}
\begin{equation}
R_{if}=R_1+R_2,
\end{equation}
\begin{equation}
R_{of}=R_1||R_2.
\end{equation}
\end{subequations}
Note that, for the circuits connected as above, the output loading computation would require $i_{if}$ to be shorted to ground. For the input loading, we note that the output is already open-circuit due to the nature of the circuit so students not need change anything.
\par
Looking at cases (a) to (d) in Fig. 3, we can see a simplified pattern. In all cases, the loading of the feedback circuit on the forward path can be easily calculated from the bottom pictures without memorizing anything.
For the output loading a short-circuit is already there in the input shunt circuits [see Figs. 3(a) and 3(d)], and an open is visible for series circuits [see Figs. 3(b) and 3(c)]. For the input loading voltage sources visible in the output shunt circuits need to be shorted, while the current sources in the output series circuits naturally need to be opened.
\par
The feedback circuit analysis using four-port models is simplified as follows. First the students must recognize the feedback types based on the rules outlined in Fig. 1 and Fig. 2. Then, the feedback circuit must be isolated and driven by voltage or current sources as required. Finally, the loading effects can be extracted by almost simple inspection without memorizing anything.
\par
After determining the type of feedback circuit and computing its loading effect, it remains to calculate the input and output impedances, which is expected to be straightforward.  In fact, a missing link in teaching feedback circuit analysis to engineering students is calculating the output impedance in the case of sensing current at output port (series feedback at the output), which is of a great consequence and is not covered in any textbook. For example, in Gray and
Meyer's book \cite{nla.cat-vn2245999} the only vaguely studied case is limited to the series-series feedback with the output assigned to the collector and the output feedback connected to the emitter, and no comment has been made on the other case, when the output is sensed at the emitter and the feedback branch is connected to the collector. We will go through this calculation during the next section.

\subsection{Derivation of the Output Impedance of Series-Series Feedback}
While the two-port approach is the most popular method of analyzing feedback loops, not all kinds of feedback circuits have been analyzed with this technique in different main references for teaching circuit analysis \cite{nla.cat-vn2245999}, \cite{12}. Referring to these textbooks, the output impedance of the output series feedback has remained unclear based on the two-port approach. We find only a vague explanation that the output impedance will increase as a result of the output series feedback for instance (section 8.5.2, in \cite{nla.cat-vn2245999} page 575). In its explanation, it has been mentioned  that the output impedance is increased by a factor of $1+T$, where $T$ is the loop gain, however what is meant by the output impedance? To enlighten this matter, let us go more thoroughly through the concept of output series feedback and its effect. What is controlled by this feedback circuit is the variation in output current which can be interpreted as an enlarged output resistance in the branch determining the changes in the output current, which we call the output branch. We represent all prior stages with a Thevenin equivalent of $v_{th}=K.v_{in}$ and $R_{th}=r_{out}$, where $K$ is the gain of the prior stages and $r_{out}$ is the output resistance of all previous stages.
\begin{figure}[]
\includegraphics[scale=0.8]{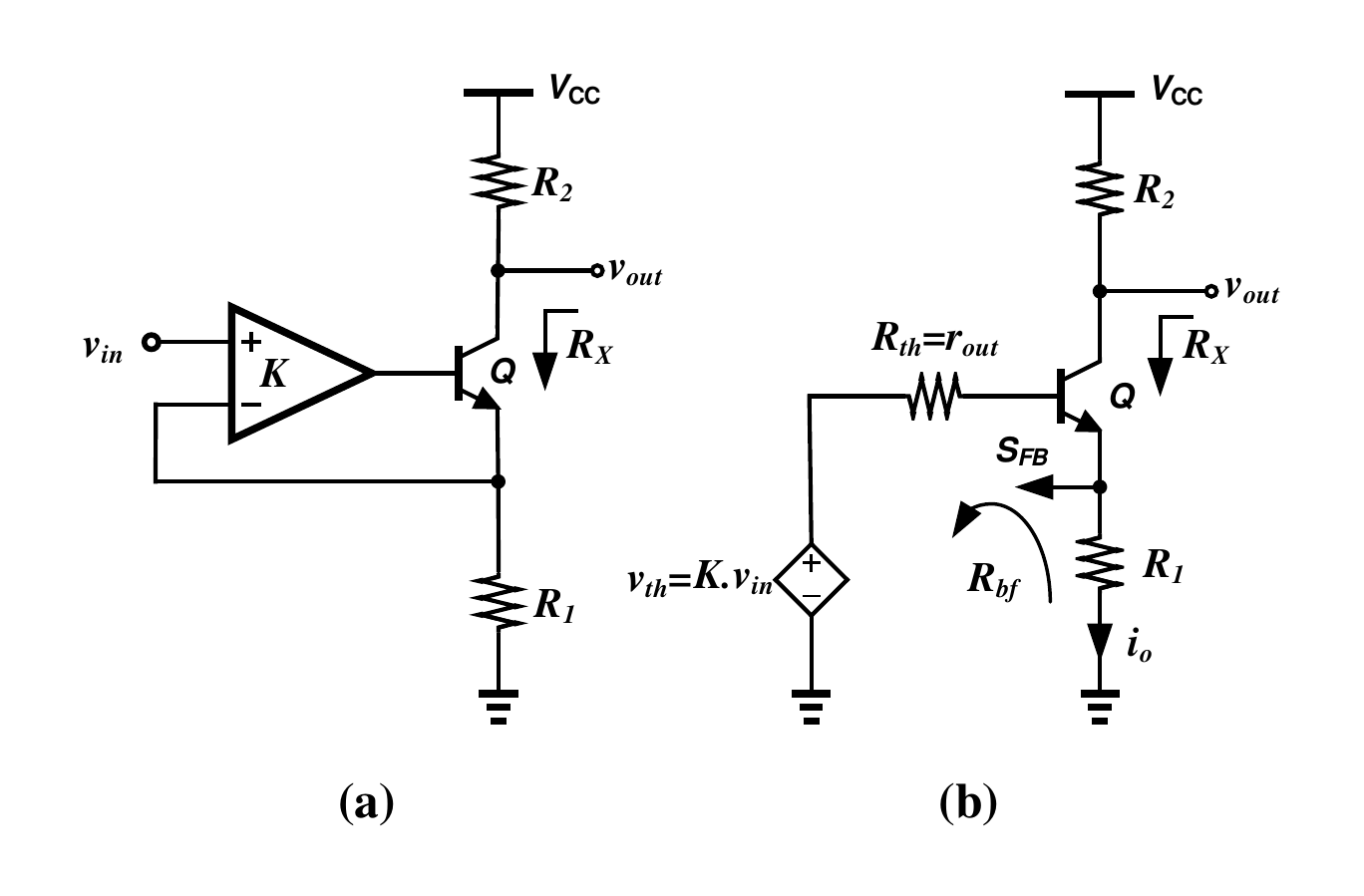}
\centering
\caption{An amplifier circuit with the output series feedback (output assigned to the collector and the current sensed through the emitter):  a) Schematic, b) Thevenin equivalent circuit.}
\label{Figure. 3}
\end{figure}
\begin{figure}[]
\includegraphics[scale=0.8]{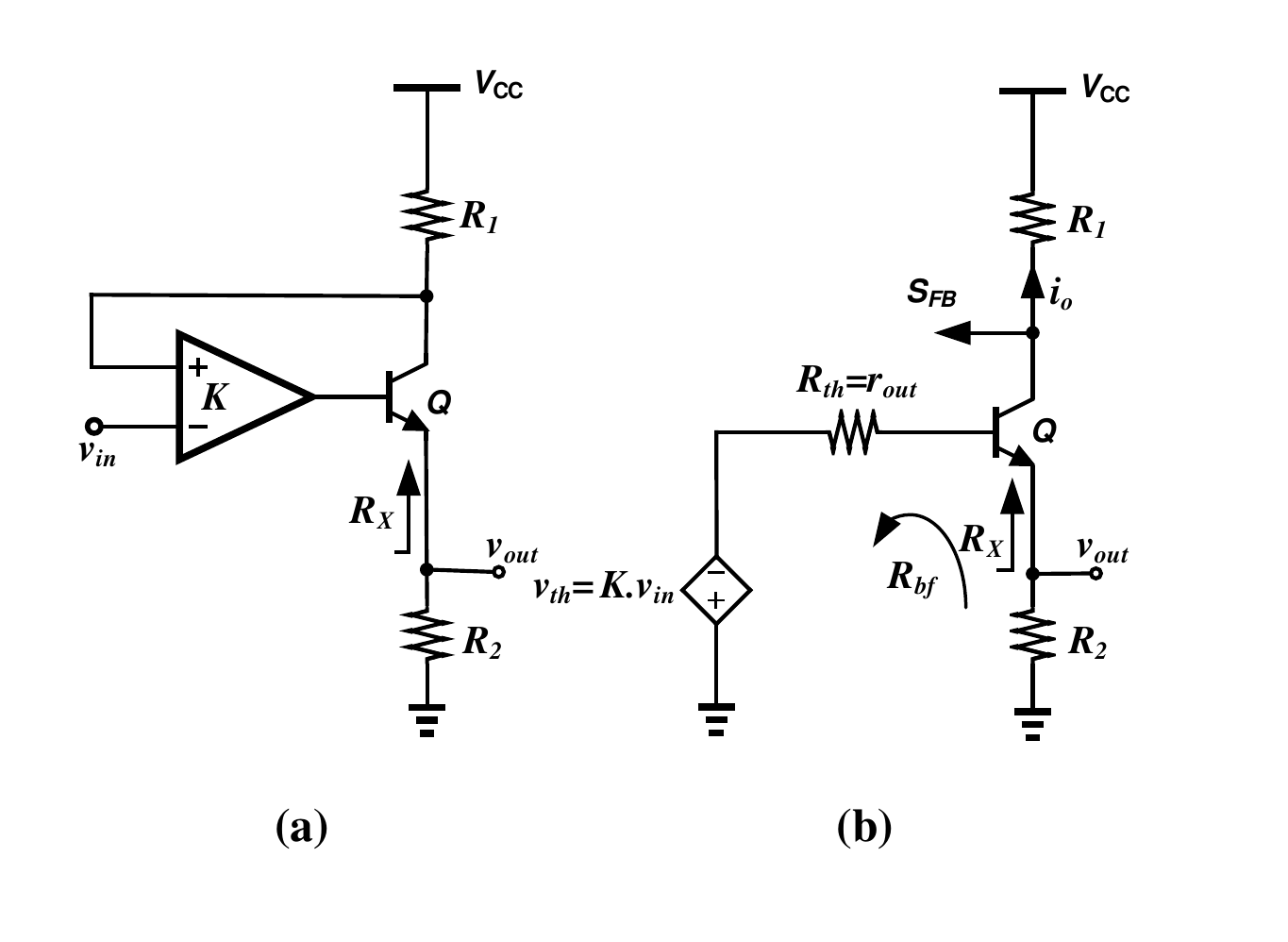}
\centering
\caption{An amplifier circuit with output series feedback (output assigned to the emitter and the current is sensed through collector):  a) Schematic b) Thevenin equivalent circuit.}
\label{Figure 4}
\end{figure}
Now, the output current for the typical equivalent circuit shown in Fig. 4(b) (neglecting the base series resistance) could be written as
\begin{equation}
i_o=\frac{K.v_{in}}{R_1+\frac{r_{out}+r_{\pi}}{\beta+1}}.
\end{equation}
\begin{figure*}[]
\includegraphics[scale=0.75]{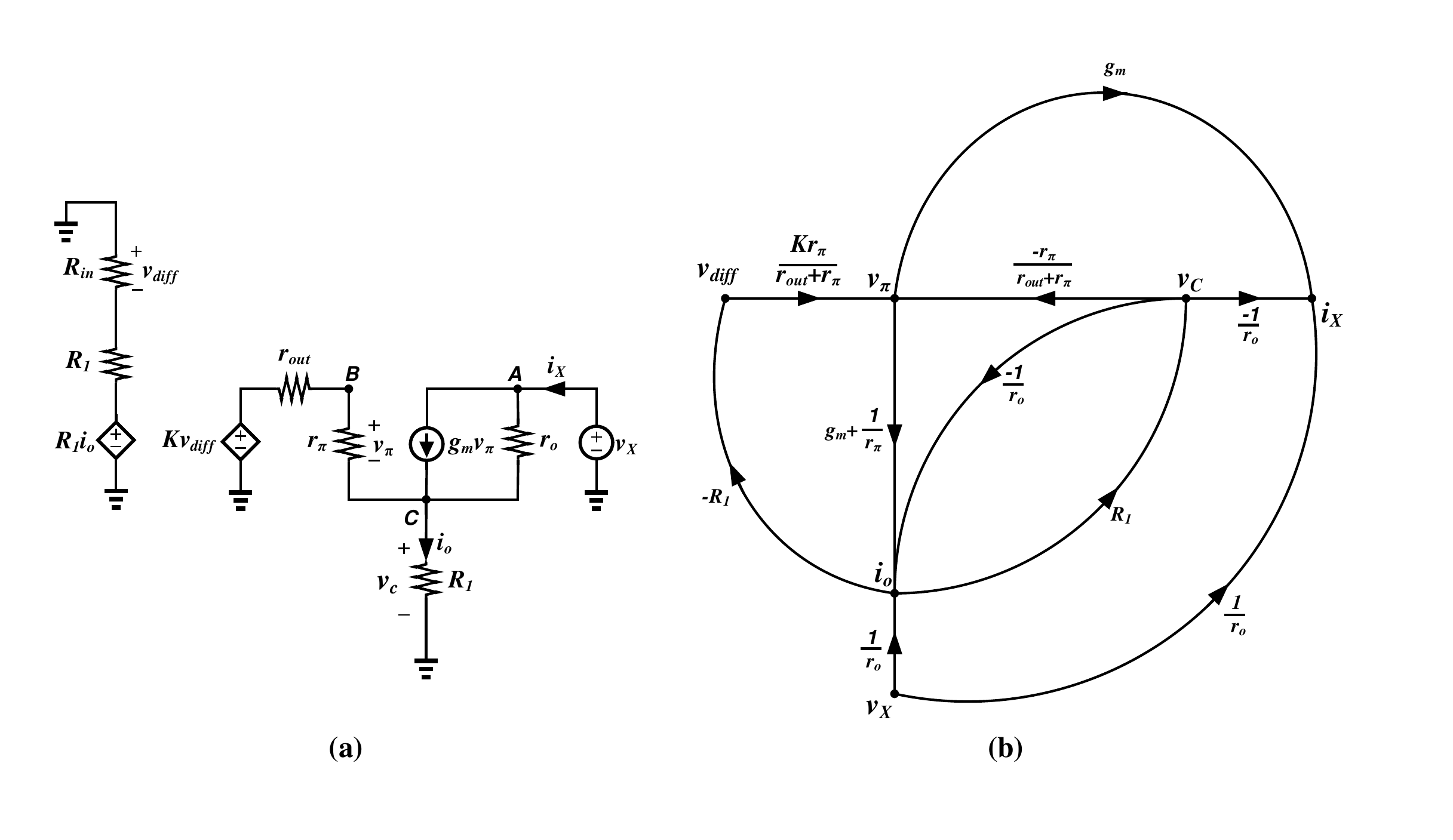}
\centering
\caption{Equivalent circuit of the first case (see Fig. 4), with the circuit replaced with its a) hybrid-$\pi$ model, b) Flow graph model.}
\label{Figure 6}
\end{figure*}
The branch containing $R_1$, $r_e=r_\pi/(\beta+1)$, and $r_{out}$, is hereafter called the output branch. Therefore, by adding the output series feedback to the circuit [see Fig. 4(b)] the output branch resistance with feedback, $R_{bf}$, will be equivalent to
\begin{equation}
R_{bf}=(R_1+\frac{r_{out}+r_{\pi}}{\beta+1})(1+af) \label{13}.
\end{equation}
As is shown in Fig. 4(a) the series feedback is connected to the emitter of the output transistor and thus the output resistance, is measured in the output transistor collector. For the open-loop circuit, the output impedance could be calculated based on the equation presented in reference \cite{nla.cat-vn2245999} for the common emitter output stage with the emitter degeneration resistance, $R_E$ and base resistance, $R_S$
\begin{equation}
R_X \simeq r_o\frac{1+g_mR_E+g_m\frac{R_S}{\beta}}{1+g_m\frac{R_E}{\beta}+g_m\frac{R_S}{\beta}}.
\end{equation}
Considering the closed-loop feedback circuit, the emitter resistance, $R_E$, is a part of the output branch and could be approximately replaced by $R_{bf}$ [from equation \eqref{13}]
\begin{equation}
R_X \simeq r_o\frac{1+g_m(1+af)(R_1+\frac{r_{out}+r_{\pi}}{\beta+1})+g_m\frac{r_{out}}{\beta}}{1+g_m\frac{(1+af)(R_1+\frac{r_{out}+r_{\pi}}{\beta+1})}{\beta}+g_m\frac{r_{out}}{\beta}} \label{666}.
\end{equation}
$1+af$ is approximately equal to $\frac{R_1(K+1)+\frac{r_{out}+r_{\pi}}{\beta+1}}{R_1+\frac{r_{out}+r_{\pi}}{\beta+1}}$, where $K$ and $r_{out}$ denote the gain and the output resistance of the op-amp, respectively, therefore we can rewrite \eqref{666} as
\begin{equation}R_X\simeq
r_o\frac{1+g_m(\frac{R_1(K+1)(\beta +1)+r_{out}+r_{\pi}}{\beta+1})+g_m\frac{r_{out}}{\beta}}{1+g_m(\frac{R_1(K+1)(\beta +1)+r_{out}+r_{\pi}}{(\beta+1)\beta})+g_m\frac{r_{out}}{\beta}}\label{11}.
\end{equation}
So we can rewrite \eqref{11} as
\begin{equation}R_X\simeq
r_o\frac{R_1(K+1)(\beta +1)+2r_{out}+2r_{\pi}}{\frac{R_1(K+1)(\beta +1)+r_{out}(\beta +1)+r_{\pi}(\beta +1)}{\beta}}\label{16}.
\end{equation}
If $R_1(\beta+1)(K+1)$ is much greater than $r_{out}(\beta+1)+r_{\pi}$, we can rewrite \eqref{11} as
\begin{equation}R_X\simeq r_o\frac{1+R_1g_m(K+1)}{1+R_1\frac{g_m(K+1)}{\beta}} \label{17}.
\end{equation}
Thus, the overall output resistance is the parallel equivalent of the value obtained from \eqref{16} and the collector resistance,  meaning:
$R_{out}=R_2||R_X $.
Generally, $ R_2 \ll R_X$ and thus the total output resistance will be close to $R_2$. This approximate value has always been used in the previous references due to the fact that usually $R_2 \ll R_X$ and there seemed no need to derive $R_X$. However in today's CMOS design, $R_2$ is normally replaced by a current source, while on the other hand the output resistance, $r_o$, of nanometer MOS devices is increasingly becoming smaller, calling for a more rigorous analysis like the above.
\par
The second model for the connection of the output series feedback which could be seen in different circuits is where the feedback network is connected to collector of the output transistors and thus the output voltage and therefore the output resistance is measured at the emitter  of this transistor [Fig. 5(a)]. As it can be inferred from this figure,
$
R_{out}=R_X||R_2
$
where $R_X$ is the resistance seen upward through the emitter. Thus, to complete the relation, we should derive the value of $R_X$. As it was mentioned in the previous section, the output branch comprises $R_2, r_e=r_\pi/\beta$, and $ r_{out}$, therefore, by adding the output series feedback to the circuit [see Fig. 5(b)] the
output branch resistance with feedback, $R_{bf}$, will be equivalent to
\begin{equation}
R_{bf}=(R_2+\frac{r_{out}+r_{\pi}}{\beta})(1+af).
\end{equation}
Furthermore, the output branch impedance could be rewritten as
\begin{equation}
R_{bf}=R_2+R_X,
\end{equation}
and therefore
\begin{equation}
R_X=R_{bf}-R_2 \label{12}.
\end{equation}
$1+af$ is equal to
$1+\frac{KR_1}{R_2+\frac{r_{out}+r_{\pi}}{\beta}}$, where $K$ and $r_{out}$ denote the gain and the output
resistance of the op-amp, respectively, so we can rewrite \eqref{12} as
\begin{equation}
R_X=\frac{R_1K \beta+r_{out}+r_\pi}{\beta}\label{14}.
\end{equation}
As a result, the overall output impedance will be
\begin{equation}
R_{out}=R_X||R_2\label{15},
\end{equation}
while replacing the value of $R_X$ from \eqref{15} yields
\begin{equation}
R_{out}=\frac{R_1K \beta+r_{out}+r_\pi}{\beta}||R_2.
\end{equation}
\par
Given that the output series feedback always occurs in one of the two forms considered here, the presented analysis could be used in any kind of output series feedback circuit to derive the output impedance easily. In order to confirm our proposed approach, in the next section, the derived relations for output resistance has been confirmed through the well-known SFG and KVL-KCL methods for a typical amplifier circuit.
\section{THE PROPOSED ANALYSIS JUSTIFICATION THROUGH SFG AND KVL-KCL ANALYSIS}
In this section, we derive the output resistance of the two discussed types of output series feedback with two well-known methods of SFG and KVL-KCL, for the sample amplifier circuits shown in Fig. 4 and Fig. 5.\par
The brief derivation of the output impedance for the first case (see Fig. 4) with SFG and KVL-KCL will be explained in the following.
\subsection{First Output Series Feedback Analysis Through SFG}
In this section, first type of output series feedback (see Fig. 4) is analyzed through SFG method. The output impedance is derived by substituting transistor's hybrid-$\pi$ model where $v_{in} = 0$ and a test voltage $(v_X)$ is used to drive
the amplifier output, and the resulting $i_X$ is then calculated [see Fig. 6(a)].\par
This model simplifies to
\begin{equation}i_o=\frac{v_\pi}{r_\pi}+g_mv_\pi +\frac{v_X-v_C}{r_o} \label{a1},
\end{equation}
in which $v_C$ and $i_X$ is given by
\begin{figure}[]
\includegraphics[scale=0.85]{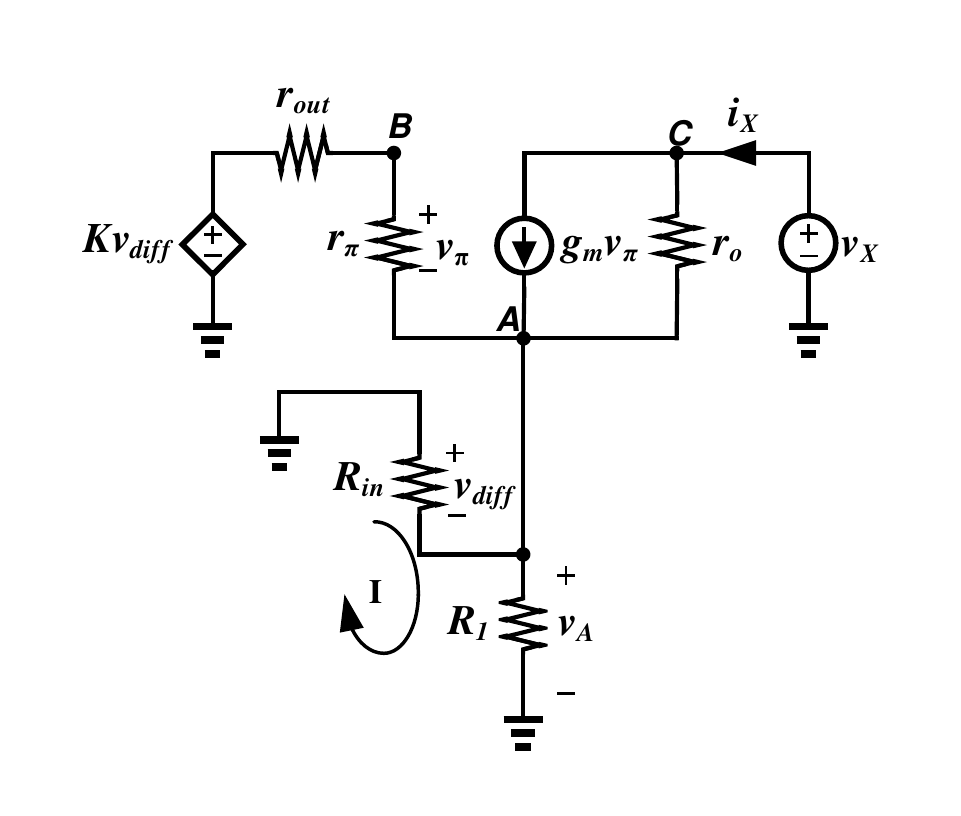}
\centering
\caption{The hybrid-$\pi$ model of the first case (see Fig. 4), for output impedance calculation.}
\label{Figure 8}
\end{figure}
\begin{figure*}[]
\includegraphics[scale=0.75]{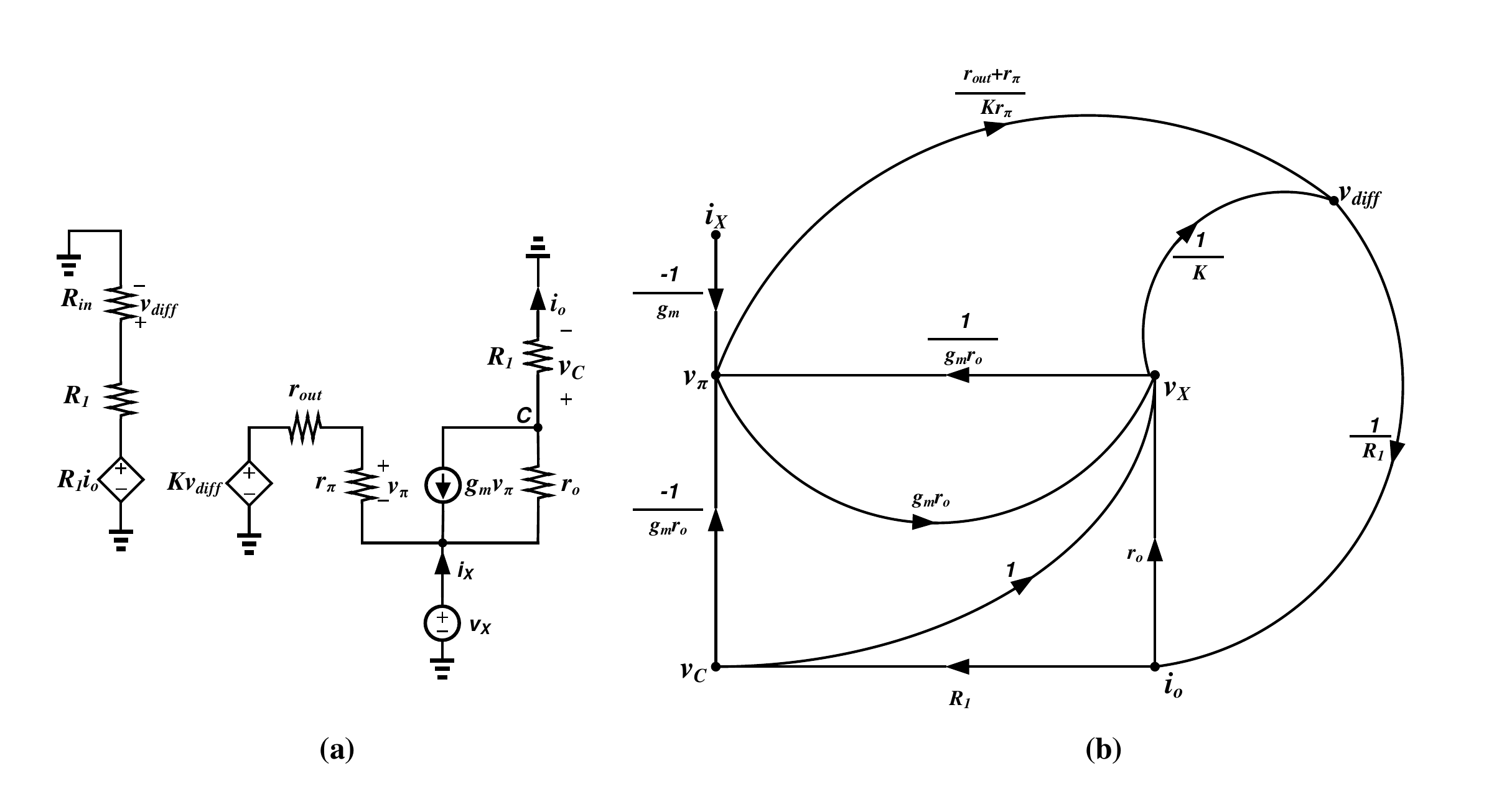}
\centering
\caption{Equivalent circuit of the second case (see Fig. 5), with the circuit replaced with its a) hybrid-$\pi$ model, b) Flow graph model.}
\label{Figure 10}
\end{figure*}
\begin{equation}v_C=i_oR_1,
\end{equation}
and
\begin{equation}i_X=\frac{v_X - v_C}{r_o}+g_mv_\pi ,
\end{equation}
respectively, where
\begin{equation}v_\pi =\frac{(Kv_{diff}-v_C)r_\pi}{r_{out}+r_\pi},
\end{equation}
and
\begin{equation}v_{diff}=-R_1i_o\frac{R_{in}}{R_{in}+R_1}.\label{x1}
\end{equation}
Hence, $R_{in} \gg R_1$,  then $R_{in}/(R_{in}+R_1)\simeq 1$.
\newline
\newline
So we can rewrite \eqref{x1} as
\begin{equation}v_{diff}=-R_1i_o.
\end{equation}
\par
These equations can be represented graphically as shown in Fig. 6(b).
According to Mason's formula \cite{6865661}, the transmission gain, $1/R_X$, from node $v_X$ to node $i_X$ is
\begin{equation}
\frac{i_X}{v_X}=\frac{1}{R_X}=\frac{1}{\Delta}\sum_{k}P_k\Delta_k \label{a2},
\end{equation}
where $P_k$ is the gain of the $k^{th}$ forward path, $\Delta$ is the determinant of the graph, and $\Delta_k$ is the determinant with the $k^{th}$ forward path eliminated.
\par
The determinant is then calculated by
\begin{align}\Delta=1-[\frac{Kr_\pi}{r_{out}+r_\pi}(g_m+\frac{1}{r_\pi})(-R_1)
\nonumber \\+R_1\frac{-1}{r_o}
+ R_1\frac{-r_\pi}{r_{out}+r_\pi}(g_m+\frac{1}{r_\pi})] \label{a3}.
\end{align}
$P_k$ and $\Delta_k$ for $k=1,2,3,4$ is as
\begin{subequations}
\begin{equation}P_1=\frac{1}{r_o},
\end{equation}
\begin{equation}
\Delta_1=\Delta,
\end{equation}
\begin{equation}P_2=\frac{1}{r_o}R_1\frac{-1}{r_o},
\end{equation}
\begin{equation}
\Delta_2=1,
\end{equation}
\begin{equation}P_3=\frac{1}{r_o}R_1\frac{-r_\pi}{r_{out}+r_\pi}g_m,
\end{equation}
\begin{equation}
\Delta_3=1,
\end{equation}
\begin{equation}P_4=\frac{1}{r_o}(-R_1)\frac{Kr_\pi}{r_{out}+r_\pi}g_m,
\end{equation}
\begin{equation}
\Delta_4=1  \label{a4}.
\end{equation}
\end{subequations}
Substituting \eqref{a3} through \eqref{a4} into \eqref{a2} and rearranging gives
\begin{equation}R_X=\frac{r_o(1+R_1\frac{(\beta +1)(K+1)}{r_{out}+r_\pi})+R_1}{1+\frac{R_1(K+1)}{r_{out}+r_\pi}}\label{5b1}.
\end{equation}
Rearranging \eqref{5b1} obtains
\begin{equation}R_X\simeq
\frac{r_o(R_1(K+1)(\beta +1)+r_{out}+r_{\pi})+R_1(r_{out}+r_\pi)}{R_1(K+1)+r_{out}+r_{\pi}}\label{a6}.
\end{equation}
If $r_\pi \gg r_{out}$ and $\frac{r_o(\beta +1)(K+1)}{r_{out}+r_\pi} \gg 1$,
we can rewrite \eqref{5b1} as
\begin{equation}R_X=r_o\frac{1+R_1g_m(K+1)}{1+R_1\frac{g_m(K+1)}{\beta}}\label{a5},
\end{equation}
which confirms the obtained equation in \eqref{17}.
\par
In order to find total output resistance, $R_X$ should be considered
in shunt with $R_2$.
\subsection{First Output Series Feedback Analysis Through KVL-KCL Analysis}
In this part, we are going to reconfirm the derived equation for output impedance in the first type of output series feedback through KVL-KCL analysis. Therefore, the circuit can be analyzed by writing the KVL-KCL nodal relations of its hybrid-$\pi$ model where $v_{in}$ = 0 and a test voltage ($v_X$) is used to drive the
amplifier output and the resulting $i_X$ is then calculated (see Fig. 7).
\par
KCL at node A gives
\begin{equation}\frac{v_A}{R_{in}} + \frac{v_A}{R_1} = \frac{v_\pi}{r_\pi} + g_mv_\pi + \frac{v_X - v_A}{r_o}\label{b1}.
\end{equation}
KCL at node B gives
\begin{equation}\frac{v_A + v_\pi - Kv_{diff}}{r_{out}} + \frac{v_\pi}{r_\pi}=0.
\end{equation}
KVL around loop I gives
\begin{equation}
v_{diff}=-v_A.
\end{equation}
KCL at node C gives
\begin{equation}i_X=\frac{v_X - v_A}{r_o} + g_mv_\pi \label{b2}.
\end{equation}
Combination of \eqref{b1} to \eqref{b2} obtains
\begin{equation}\frac{v_X}{i_X}=R_X=\frac{r_o(1+R_1\frac{(\beta +1)(K+1)}{r_{out}+r_\pi})+R_1}{1+\frac{R_1(K+1)}{r_{out}+r_\pi}}\label{b4}.
\end{equation}
Rearranging \eqref{b4} obtains
\begin{equation}R_X\simeq
\frac{r_o(R_1(K+1)(\beta +1)+r_{out}+r_{\pi})+R_1(r_{out}+r_\pi)}{R_1(K+1)+r_{out}+r_{\pi}}\label{b5}.
\end{equation}
If $r_\pi \gg r_{out}$ and $\frac{r_o(\beta +1)(K+1)}{r_{out}+r_\pi} \gg 1$,
we can rewrite \eqref{b4} as
\begin{equation}R_X\simeq r_o\frac{1+R_1g_m(K+1)}{1+R_1\frac{g_m(K+1)}{\beta}} \label{b3},
\end{equation}
which again confirms \eqref{17}.
\par
To find total output resistance, $R_X$ should be considered
in shunt with $R_2$.
\par
For typical values of the circuit parameters, \textit{i.e.,} $\beta=100$, $K=1000$, $r_{out}=500k\Omega$, $R_1=1k\Omega$, $r_\pi=2.5k\Omega$, $r_{o}=100K\Omega$, equation \eqref{16} provides an output impedance, $R_X=6.724M\Omega$ while equations \eqref{a6} and \eqref{b5} give $R_X=6.758M\Omega$, that results in an error of $0.5\%$.
\par
In a similar manner, the output impedance for the second type of output series feedback (see Fig. 5) is also verified with both SFG and KVL-KCL analysis while the corresponding equations are briefly described in the next two subsections.
\subsection{Second Output Series Feedback Analysis Through SFG}
In this subsection, the second type of output series feedback circuit is analyzed by substituting the transistor's hybrid-$\pi$ model where $v_{in} = 0$ and a test voltage $(v_X)$ is used to drive
the amplifier output, and the resulting $i_X$ is then calculated [see Fig. 8(a)]. The corresponding circuit equations are briefly presented here.
\begin{figure}[]
\includegraphics[scale=0.85]{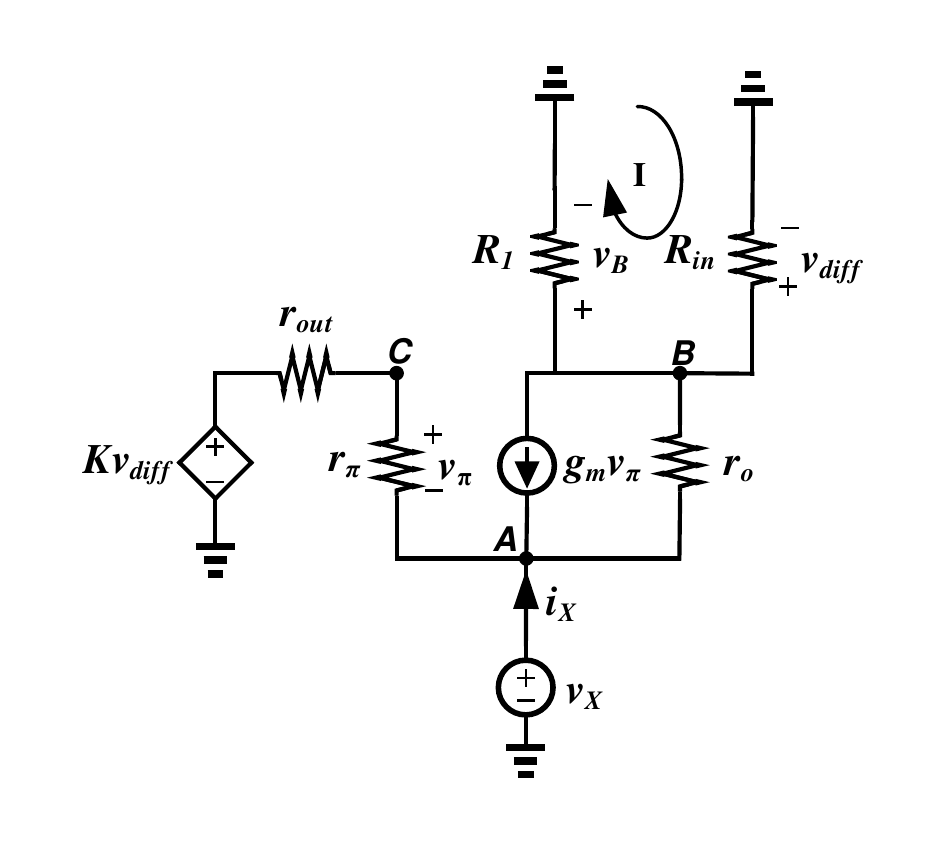}
\centering
\caption{The hybrid-$\pi$ model of the second case (see Fig. 5), for output impedance calculation.}
\label{Figure11}
\end{figure}
\par
This model simplifies to
\begin{equation}
i_X=\frac{-v_\pi}{r_\pi}-g_mv_\pi -\frac{v_C-v_X}{r_o}\label{c1}.
\end{equation}
As $g_m+\frac{1}{r_\pi}\simeq g_m$, rearranging \eqref{c1} gives
\begin{equation}
v_\pi =\frac{-i_X}{g_m}+\frac{-v_C}{g_mr_o}+\frac{v_X}{g_mr_o}\label{c2}.
\end{equation}
$v_C$ in \eqref{c2} can be derived
as
\begin{equation}
v_C=i_oR_1.
\end{equation}
The test voltage ($v_X$) is given by
\begin{equation}
v_X=i_or_o+v_\pi g_mr_o+v_C,
\end{equation}
where
\begin{equation}
v_{diff}=\frac{v_\pi(r_{out}+r_\pi)}{Kr_\pi}+\frac{v_X}{K},
\end{equation}
and
\begin{equation}
R_1i_o\frac{R_{in}}{R_{in}+R_1}=v_{diff}\label{3b2}.
\end{equation}
Since $R_{in} \gg R_1$, we can rewrite \eqref{3b2}
\begin{equation}
i_o=\frac{v_{diff}}{R_1}.
\end{equation}
These equations can be represented graphically as shown in Fig. 8(b).
According to Mason's formula, the transmission gain, $R_X$, from node $i_X$ to node $v_X$ is
\begin{equation}
\frac{v_X}{i_X}=R_X=\frac{1}{\Delta}\sum_{k}P_k\Delta_k \label{c3},
\end{equation}
where $P_k$ is the gain of the $k^{th}$ forward path, $\Delta$ is the determinant of the graph, and $\Delta_k$ is the determinant with the $k^{th}$ forward path eliminated.
\par
The determinant can be calculated by
\begin{equation}
\Delta =-\frac{r_o\beta +r_{out}+r_\pi}{R_1K\beta}\label{c4}.
\end{equation}
$P_k$ and $\Delta_k$ for $k=1,2,3$ is as
\begin{subequations}
\begin{equation}
P_1=\frac{-1}{g_m}\frac{r_{out}+r_\pi}{Kr_\pi}\frac{1}{R_1}r_o,
\end{equation}
\begin{equation}
\Delta_1 =1,
\end{equation}
\begin{equation}
P_2=\frac{-1}{g_m}\frac{r_{out}+r_\pi}{Kr_\pi}\frac{1}{R_1}R_11,
\end{equation}
\begin{equation}
\Delta_2 =1,
\end{equation}
\begin{equation}
P_3=\frac{-1}{g_m}g_mr_o,
\end{equation}
\begin{equation}
\Delta_3 =1\label{c5}.
\end{equation}
\end{subequations}
Substituting \eqref{c4} through \eqref{c5} into \eqref{c3} and rearranging gives
\begin{equation}R_X=\frac{r_oR_1K\beta+r_o(r_{out}+r_\pi)+R_1(r_{out}+r_\pi)}{r_o\beta +r_{out}+r_\pi} \label{c7}.
\end{equation}
Since $r_o\beta \gg r_{out}+r_\pi$ and $r_o \gg R_1$,
\newline
\newline
we can rewrite \eqref{c7} as
\begin{equation}R_X=\frac{R_1K\beta +r_\pi + r_{out}}{\beta}\label{c6},
\end{equation}
which is exactly similar to the derived equation for $R_X$ in \eqref{14}.
\par
In order to find total output resistance, $R_X$ should be considered
in shunt with $R_2$.
\subsection{Second Output Series Feedback Analysis Through KVL-KCL Analysis}
Finally, the derived  output resistance for the second type of output series feedback circuit is approved here through KVL-KCL equations. For this calculation, consider the equivalent circuit shown in
Fig. 9, where $v_{in}$ = 0 and a test voltage ($v_X$) is used to drive the
amplifier output and the resulting $i_X$ is then calculated.
\par
KCL at node A gives
\begin{equation}
i_X + g_mv_\pi + \frac{v_B - v_X}{r_o} + \frac{v_\pi}{r_\pi}=0\label{d1}.
\end{equation}
KCL at node B gives
\begin{equation}
\frac{v_B}{R_1||R_{in}} + \frac{v_B - v_X}{r_o} + g_mv_\pi=0.
\end{equation}
KVL around loop I gives
\begin{equation}
v_{diff}=v_B.
\end{equation}
KCL at node C gives
\begin{equation}
\frac{v_\pi}{r_\pi}=\frac{Kv_{diff} - (v_X + v_\pi)}{r_{out}}\label{d2}.
\end{equation}
Combination of \eqref{d1} to \eqref{d2} obtains
\begin{equation}R_X=\frac{r_oR_1K\beta+r_o(r_{out}+r_\pi)+R_1(r_{out}+r_\pi)}{r_o\beta +r_{out}+r_\pi} \label{d4}.
\end{equation}
Since $r_o\beta \gg r_{out}+r_\pi$ and $r_o \gg R_1$,
\newline
\newline
we can rewrite \eqref{d4} as
\begin{equation}
\frac{v_X}{i_X}=R_X=\frac{R_1K\beta +r_\pi + r_{out}}{\beta}\label{d3}.
\end{equation}
Again this equation is in good agreement with \eqref{14}.
\par
To find total output resistance, $R_X$ should be considered
in shunt with $R_2$.
\par
For typical values of the circuit parameters, \textit{i.e.,} $\beta=100$, $K=1000$, $r_{out}=500k\Omega$, $R_1=1k\Omega$, $r_\pi=2.5k\Omega$, $r_{o}=100k\Omega$, equation \eqref{14} provides an output impedance, $R_X=1.005M\Omega$ and equations \eqref{c7} and \eqref{d4} give $R_X=0.956M\Omega$ that results in an error of $5.01\%$.
\par
The main contributors to this error are the op-amp's  non idealities. For instance, the output resistance is assumed to be $500k\Omega$, whereas in a well designed op-amp in which  a proper output stage is utilized, this value will be as small as $10\Omega$, resulting in a negligible error.
\par

\section{DISCUSSION AND CONCLUSIONS}
This paper demonstrated a simple method of deciding the feedback types followed by a more intuitive two-port analysis. This analysis covers all types of possible feedback circuits. Furthermore, in the second part of this paper, a simplified approach to the analysis of the ambiguous series feedback at the output was presented and rigorously proven via the well-known SFG and KVL-KCL methods.\par
The new method was used in class for more than six years. The methodology was shown on 20 circuit examples on the author's class website.
\footnote{http://ee.sharif.edu/$\sim$elecprinc-SatMon.} The same circuits were also analyzed using Blac
kmans's method \cite{6767497} and our experience shows that the proposed feedback type recognition method removed many ambiguities that students confront when identifying the feedback box in the two-port analysis. The issue of determining the output impedance in series feedback at the output also removes ambiguities.


\newpage


%


\ifCLASSOPTIONcaptionsoff
  \newpage
\fi



\bibliographystyle{IEEEtran}
\bibliography{IEEEabrv,Bibliography}


%



\vspace{-10mm}
\begin{IEEEbiography}[{\includegraphics[width=1in,height=1.25in,clip,keepaspectratio]{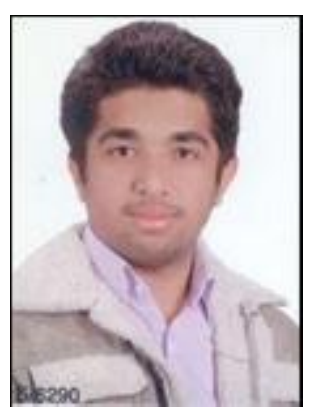}}]{Morteza  Fayazi} was born in 1994, Tehran, Iran.  He received the B.S. degree in electrical engineering from Sharif University of Technology in 2017, and M.S. degree in computer science from University of Michigan in 2019. He is currently working toward his PhD degree at department of electrical engineering and computer science at University of Michigan.
\end{IEEEbiography}
\vspace{-20mm}
\begin{IEEEbiography}[{\includegraphics[width=1in,height=1.25in,clip,keepaspectratio]{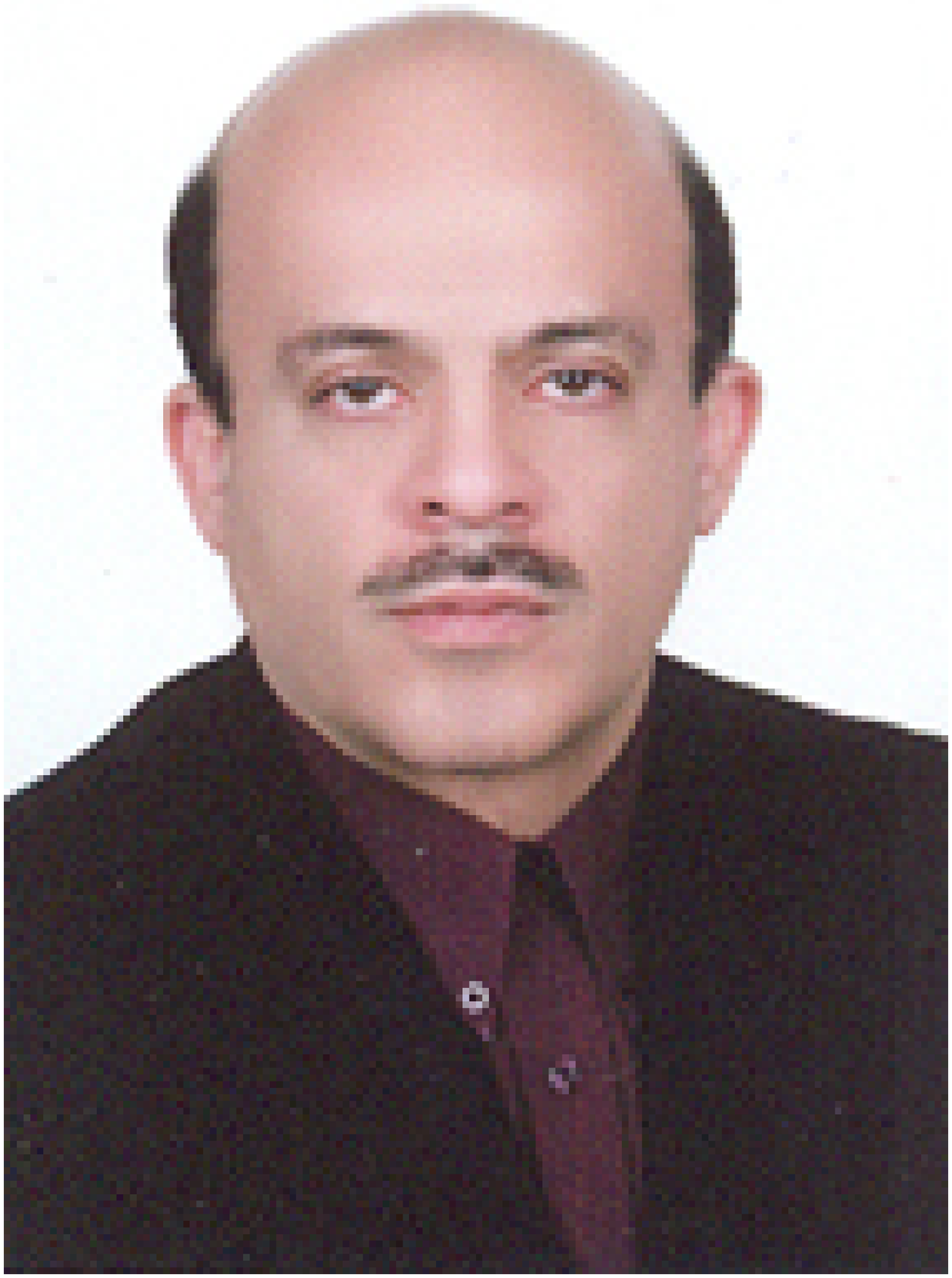}}]{Ali Fotowat-Ahmady}
(M'80) was born in Tehran,
Iran in 1958. He received the B.S. degree from Cali-
fornia Institute of Technology in 1980, and M.S. and
PhD degrees in electrical engineering from Stanford
University in 1982 and 1991 respectively.
He started his career at Philips Semiconductor in
Sunnyvale, California in 1987 where he developed
several integrated circuits for mobile phones. In
1991 he joined Sharif University of Technology
EE Department where he is an Associate Professor. He is a three times recipient of
Kharazmi science and engineering award for his
work on low power microelectronics and communication ICs. His research
interests include advanced integrated circuits for energy savings and communication/positioning applications. Due to his interests in entrepreneurial
engineering he has been the co-founder of several companies and continues
advising his students on the same. He is a member of the IEEE Solid State
Society and has been the adviser of the societies Sharif EE student chapter.
\end{IEEEbiography}
\vspace{-150mm}
\begin{IEEEbiography}[{\includegraphics[width=1in,height=1.25in,clip,keepaspectratio]{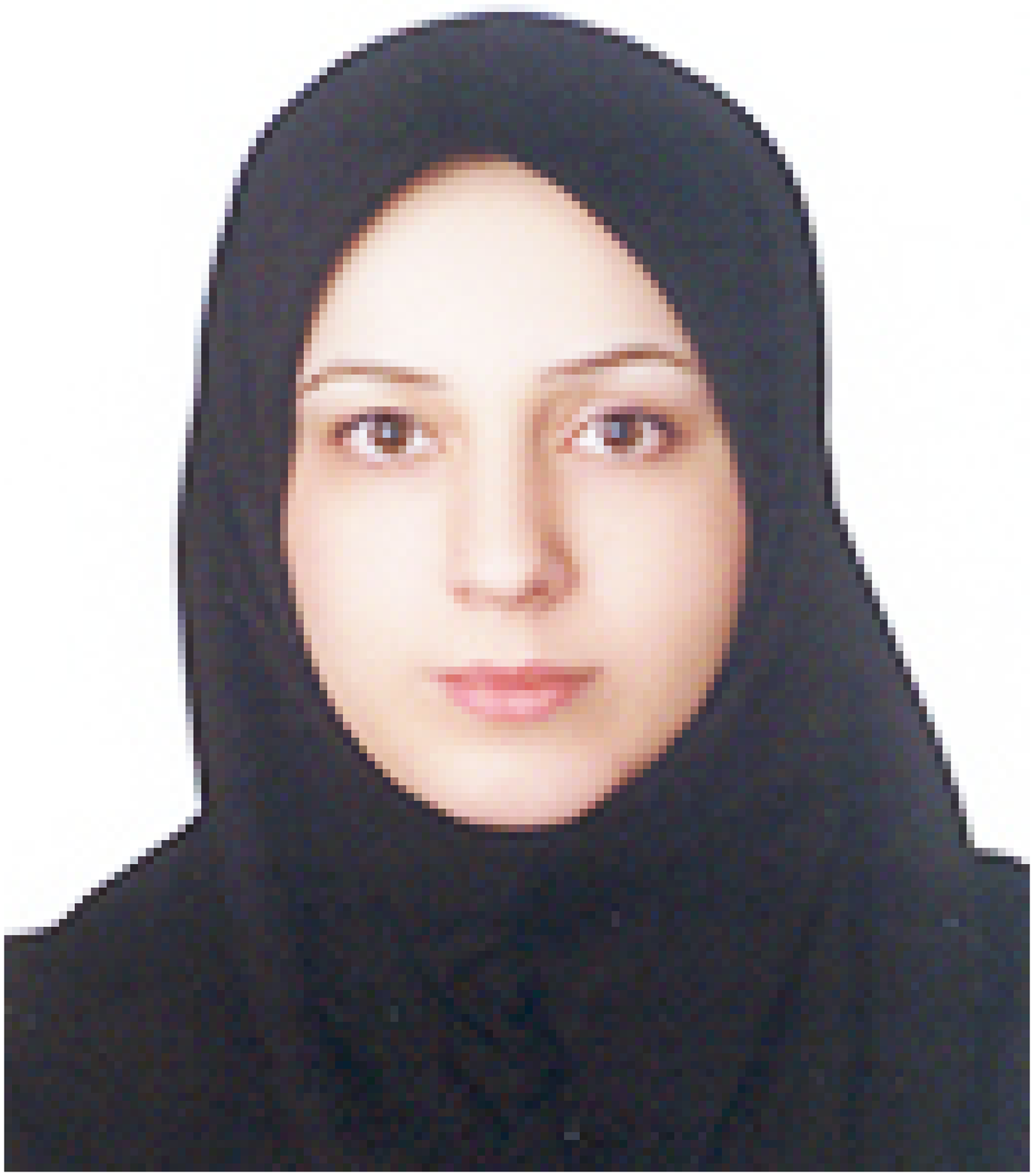}}]{Zahra Kavehvash} was born in Kermanshah, Iran, in 1983. She received the B.Sc., M.Sc. and Ph.D. degrees all in electrical engineering from Sharif University of Technology (SUT), Tehran, Iran,  in 2005, 2007 and 2012 respectively.  She joined Sharif University of Technology, EE department in 2013 as a faculty member. Her research interests include optical and millimeter
wave imaging devices, three-dimensional imaging systems, Biomedical imaging systems and optical signal processing.
\end{IEEEbiography}
\vspace{-20mm}
\end{document}